\documentclass[fleqn,twoside]{article}
\usepackage{espcrc2}
\usepackage{epsfig}

\usepackage{graphicx}
\usepackage[figuresright]{rotating}

\newcommand{\AmS}{{\protect\the\textfont2
  A\kern-.1667em\lower.5ex\hbox{M}\kern-.125emS}}
\newcommand{\Tr}{\mbox{Tr}\,}

\title{The Chiral Critical Point in 3-Flavour QCD \thanks{The work has been
    supported by the TMR network ERBFMRX-CT-970122 and by the DFG under grant
    FOR 339/1-2.}
}

\author{Ch. Schmidt\address{Fakult\"at f\"ur Physik, Universit\"at Bielefeld,
        D-33615 Bielefeld, Germany},
        F. Karsch\addressmark \,and E. Laermann\addressmark}

\begin{document}

\begin{abstract}
We determine the second order endpoint of the line of
first order phase transitions, which occur in the light quark mass
regime of 3-flavour QCD at finite temperature, and analyze
universal properties of this chiral critical point.
A detailed analysis of Binder cumulants and the joint
probability distributions of energy like and ordering-field like
observables confirms that the chiral critical point belongs to the
universality class of the 3d Ising model.
From a calculation with improved gauge and staggered fermion actions
we estimate that the transition is first order for pseudo-scalar meson
masses less than about 200 MeV.
\vspace{1pc}
\end{abstract}

\maketitle

\section{INTRODUCTION}
The phase transition in 3-flavour QCD is first order in the chiral limit
($m\equiv0$). This first order transition will continue to
persist for small but non-zero values of $m$ up to a critical value, $\bar{m}$,
of the quark mass. It is expected that QCD at this chiral critical point belongs
to the universality class of the 3d Ising model \cite{Gav94}. We will present
here evidence for this scenario \cite{schmidt}.

The lattice formulation of QCD with three degenerate quark masses
depends on the bare quark mass $m$ and the gauge coupling $\beta \equiv 6/g^2$.
The thermodynamics is described in terms of the partition function $Z(\beta, m)
= \int {\cal D}U e^{-S(\{ U\}, \beta, m)}$. The Euclidean action, $S$, is given
in terms of the pure gauge ($S_G$) and the fermion action ($S_F$)
\begin{equation}
S(\{ U\}, \beta, m) = \beta S_G(\{ U\}) - S_F(\{ U\}, m) \, ,
\label{action}
\end{equation}
where $S_F=0.75\, \Tr \ln M$ is expressed in terms of the fermion matrix $M(\{U\})$.
We will mainly use standard staggered fermions and the standard Wilson gauge
actions.

The universal properties at the chiral critical point
are controlled by an effective Hamiltonian,
\begin{equation}
{\cal H}_{eff} (\tau, \xi)  = \tau {\cal E} + \xi {\cal M} \,,
\label{heff}
\end{equation}
with a global $Z(2)$ symmetry.
It controls the critical behaviour at $(\beta_c(\bar{m}),\bar{m})$ and can be
expressed in terms of two operators $\cal E,~M$, {\it i.e.} the energy-like and
ordering-field like operators that couple to two relevant scaling
fields $\tau$ and $\xi$. As the symmetry of ${\cal H}_{eff}$ is not shared in
any obvious way by the QCD Lagrangian one may expect that in the vicinity
of the chiral critical point the operators and couplings appearing in the QCD
Lagrangian are linear combinations of ${\cal E}$, ${\cal M}$ and
$\tau$, $\xi$ respectively. One may use a linear ansatz
\begin{eqnarray}
\tau &=& \beta - \beta_c +A\; (m - \bar{m}) \, ,\nonumber \\
\xi &=& m - \bar{m} + B\; (\beta - \beta_c) \, ,
\label{tauxi}
\end{eqnarray}
as well as
\begin{equation}
{\cal E} = S_G + r\; \bar{\psi} \psi \,, \qquad
{\cal M} = \bar{\psi} \psi + s \; S_G \, .
\label{EM}
\end{equation}
Here $\bar{\psi} \psi \equiv 0.75 \, \Tr M^{-1}$ denotes the chiral condensate
expressed in terms of bosonic observables. As $S_G$, $\bar{\psi}
\psi$ or related observables like the Polyakov loop $L$ are mixtures of $\cal E$
and $\cal M$, the corresponding susceptibilities will all receive
contributions from fluctuations of $\cal E$ as well as $\cal M$.
Asymptotically therefore all of them will
show identical finite size scaling behaviour which will be dominated
by the largest universal exponent $\gamma /\nu$. Unfortunately this is quite similar
for the 3d-$Z(2)$ and $O(N)$ spin models, so that a finite size scaling
analysis of susceptibilities is not a good indicator for the universality
class. The situation is different for cumulants constructed from linear
combinations of $\bar{\psi}\psi$ and $S_G$,
\begin{equation}
B_4(x)  =  {\langle \bigl( \delta  M (x)\bigr) ^4\rangle \over
\langle \bigl( \delta  M (x)\bigr) ^2\rangle^2 } \,, \;
M(x) = \bar{\psi}\psi + x\; S_G \,.
\label{binder}
\end{equation}
Here $\delta X$ denotes $X-\langle X \rangle$. For all values of
$x$ different from $1/r$ the cumulants behave asymptotically like the
Binder cumulant for the order parameter; cumulants calculated on
different size lattices for different quark masses will intersect at some
value of the quark mass. In the infinite volume limit these intersection
points will converge to a universal value which
is characteristic for the universality class, in fact, it is quite different for
the classes of 3d-$Z(2)$ and $O(N)$ symmetric spin models;
e.g. $B_4 = 1.604$ for $Z(2)$ \cite{b4ising}, 1.242(2) for $O(2)$ \cite{b4o2}
and 1.092(3) for $O(4)$ \cite{b4o4}.
The cumulants $B_4(x)$ thus seem to be appropriate observables to locate
the chiral critical point as well as to determine its universality
class without knowing in detail the correct scaling fields.

\section{LOCATING THE CRITICAL POINT}

\begin{figure}
\begin{center}
\epsfig{file= 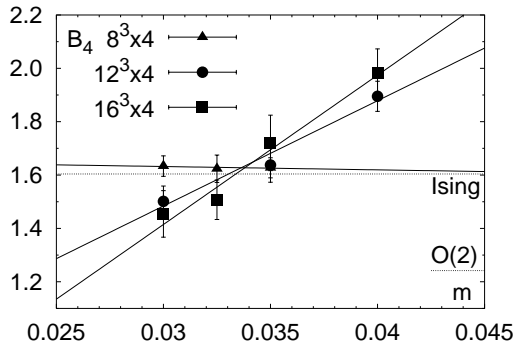,width=70mm}
\end{center}
\vspace*{-1.5cm}\caption{The cumulant $B_4 (0)$ defined in Eq. 5.}\vspace*{-0.5cm}
\label{fig:binder}
\end{figure}

We make use of the finite size scaling properties of Binder cumulants
$B_4(x)$ evaluated at $\beta_{pc}(m)$. From previous
studies one knows that the endpoint in 3-flavour QCD is located close to
$m=0.035$ \cite{Aoki99}. We have performed calculations on lattices of size
$N_\sigma^3\times 4$, with $N_\sigma = 8$, 12 and 16. We have used
four values of  the quark mass in the interval $m \in [0.03, 0.04]$
and for each of these masses we calculated thermodynamic observables for
3 to 4 different values of the gauge coupling $\beta$.
In general we collected for each pair of couplings $(1-3)\cdot 10^4$
configurations generated with the hybrid-R
algorithm\footnote{We used trajectories of length $\tau = 0.675$ generated
with a discrete step size $\delta \tau = 0.015$.}.
Interpolations between results from different $\beta$-values have been
performed using the Ferrenberg-Swendsen multi-histogram technique. The cumulant
of the chiral condensate, $B_4(0)$, is shown in Fig.~\ref{fig:binder}. We note
that the cumulants calculated on different size lattices intersect at a quark
mass close to $m=0.035$. The  value of $B_4(0)$ at the intersection point is
compatible with the universal value of the Binder cumulant for the
3d Ising model. We have fitted the cumulants on
a given lattice size using a linear ansatz in the quark mass. From this we find
for the critical mass and the cumulant,
\begin{equation}
\bar{m} = 0.033(1) \, , \quad B_4 (0)=1.639(24) \, .
\label{mcest}
\end{equation}
This result strongly suggests that the chiral
critical point indeed belongs to the universality class of the 3d
Ising model.
As discussed above the intersection point of Binder cumulants
constructed from $M(x)$ according to Eq.~\ref{binder}
will be independent of the choice of $x$ in the infinite volume limit.
The optimal choice $x=s$, will minimize the finite volume effects, which are
quadratic in $\Delta = x-s$. The determination of the $x$-value for which
the intersection point of the Binder cumulant is closest to the $Z(2)$ value
provides a finite volume estimate for the mixing parameter $s$.
One finds $s_{\rm min} = 0.430(23)$.

\section{THE MIXING PARAMETERS}

\begin{figure}
\vspace*{-1.8cm}\hspace*{-0.5cm}\epsfig{file=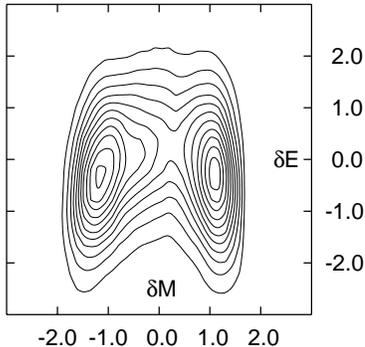,width=80mm}
\vspace*{-3.2cm}\caption{Joint probability distribution of the fluctuations in the
ordering-field like and energy like operators, which have been constructed
according to Eq.~\ref{EM} with $r=0.55$ and $s=0.43$.}\vspace*{-0.5cm}
\label{fig:jointEM}
\end{figure}

The parameters $r$ and $s$ are fixed by demanding that $\cal M$
should obey basic properties of the order parameter for
spontaneous symmetry breaking at the critical point.
We obtain two conditions for the mixing parameters $r$, $s$ in terms of
the parameter $B$ and expectation values of $S_G$, $\bar{\psi}\psi$,
\begin{eqnarray}
r &=& -B \,, \nonumber \\
s &=& {\langle \delta \bar{\psi}\psi \delta S_G \rangle -B
\langle (\delta \bar{\psi}\psi )^2 \rangle \over
\langle (\delta S_G )^2 \rangle - B \langle \delta
\bar{\psi}\psi\;  \delta S_G \rangle } \,.
\label{rs}
\end{eqnarray}
The parameter $B$ is needed to define lines of constant $\xi$.
The line of first order phase transitions defines
the zero external field line ($\xi = 0$) of the effective Hamiltonian.
We thus can extract $B$ from the quark mass dependence of the
pseudo-critical couplings. Using Eq.~\ref{tauxi} one obtains
\begin{equation}
B^{-1} = - \left.\left( {{\rm d}\beta_c(m) \over
{\rm d} m}\right)\right|_{ \; m=\bar{m}} \, .
\label{slopeB}
\end{equation}
Knowing $B$ we also know $r$ and can construct $s$ using Eq.~\ref{rs}.
A first estimate may be given using our data on the largest lattice ($16^3
\times 4$). The slope of $\beta_{pc}(m)$ defines $r^{-1}$. We estimate $r =
0.51 (2)$ from a straight line fit. The mixing parameter $r$ is large and
definitely non-zero. The importance of choosing the correct mixing parameter
$r$ becomes apparent from an analysis of joint probability distributions for
$\delta {\cal E}$ and ${\delta \cal M}$. These are little affected by changes
of $s$ but they strongly depend on $r$. In order to improve our determination
of $r$, we demand that fluctuations in ${\cal M}$ vanish for any fixed
value of $\delta {\cal   E}$. This maximizes the $Z(2)$ symmetry of the contour
plot shown in Fig.\ref{fig:jointEM}.
One obtains in this way $r=0.550(7)$. A determination of the parameter $s$
from Eq.~\ref{rs} then yields $s=0.41(51)$. Although this value
has large errors it is consistent with the result found from the
intersection point of the Binder cumulant. As a best estimate we therefore obtain
\begin{equation}
r\; =\; 0.550\pm 0.007 \,, \quad s \; =\; 0.430\pm 0.023 \,.
\label{bestmix}
\end{equation}

\section{THE PHYSICAL SCALE}

In order to determine a physical scale for the chiral critical point we have calculated
hadron masses on a $16^4$ lattice at $(\beta_c, \bar{m})$. For
the pseudo-scalar meson mass we find $m_{\rm ps}/T_c = 1.853(1)$.
Using estimates for the critical temperature \cite{Kar00} we thus estimate for
the pseudo-scalar meson mass at the chiral critical point $m_{\rm ps} \simeq
290$~MeV. The entire analysis of the  critical point discussed so far
has been performed with unimproved gauge and fermion actions on
rather coarse lattices. Improved actions are not expected to
modify the results on the universal properties of the
 critical point. They may, however,
well influence the quantitative determination of the
critical point. We therefore have investigated the  critical point also
in calculations with improved gauge and staggered fermion actions, the
p4-action \cite{Kar00,Pei99}. We have performed calculations on a lattice
of size $12^3\times 4$ with a bare quark mass $m=0.005$
and on a lattice of size $16^3\times 4$ with $m=0.01$. Again a
Ferrenberg-Swendsen reweighting is used to determine the pseudo-critical
couplings and Binder cumulants at these couplings. The smaller quark mass leads
to a Binder cumulant below the $Z(2)$ value ($B_4(0)=1.31(12)$) and for the larger quark
mass, the Binder cumulant is significantly above the $Z(2)$ value
($B_4(0)=2.14(10)$). We thus have obtained an upper and
lower limit for the  critical point, suggesting a
critical bare quark mass of $\bar{m}=0.0075(25)$.
Extrapolating the meson masses calculated in \cite{Kar00} to this
value of the bare quark mass, we estimate for the pseudo-scalar meson
mass at the  critical point $m_{\rm ps} \simeq 192(25)$~MeV.
The critical mass thus is considerably smaller than estimated from
calculations with unimproved gauge and fermion actions.

\end{document}